\documentclass[twocolumn,showpacs,preprintnumbers]{revtex4}
\usepackage{graphicx}% Include figure files
\usepackage{dcolumn}% Align table columns on decimal point
\usepackage{bm}% bold math
\usepackage[bookmarks=false]{hyperref}

\newcommand{\BoldVec}[1]{\mathchoice%
  {\mbox{\boldmath $\displaystyle     #1$}}%
  {\mbox{\boldmath $\textstyle        #1$}}%
  {\mbox{\boldmath $\scriptstyle      #1$}}%
  {\mbox{\boldmath $\scriptscriptstyle#1$}}%
}
\newcommand{\EQ}{\begin{equation}}
\newcommand{\EN}{\end{equation}}
\newcommand{\EQA}{\begin{eqnarray}}
\newcommand{\ENA}{\end{eqnarray}}
\newcommand{\eq}[1]{(\ref{#1})}

\newcommand{\Eq}[1]{Eq.~(\ref{#1})}

\newcommand{\Sec}[1]{Sec.~\ref{#1}}

\newcommand{\Fig}[1]{Fig.~\ref{#1}}

\newcommand{\Tab}[1]{Table~\ref{#1}}
\newcommand{\Figs}[2]{Figures~\ref{#1} and \ref{#2}}

\newcommand{\bra}[1]{\langle #1\rangle}

\newcommand{\xx}{\BoldVec{x}{}}

\newcommand{\uu}{\BoldVec{u} {}}

\newcommand{\BB}{\BoldVec{B} {}}

\newcommand{\AAA}{\BoldVec{A} {}}

\newcommand{\eee}{\BoldVec{e} {}}

\newcommand{\JJ}{\BoldVec{J} {}}

\newcommand{\ff}{\BoldVec{f} {}}
\newcommand{\EE}{\BoldVec{E} {}}
\newcommand{\FF}{\BoldVec{F} {}}

\newcommand{\kk}{\BoldVec{k} {}}

\newcommand{\nab}{\BoldVec{\nabla} {}}

\newcommand{\SSSS}{\mbox{\boldmath ${\sf S}$} {}}

\newcommand{\ii}{{\rm i}}

\newcommand{\DD}{{\rm D} {}}

\newcommand{\const}{{\rm const}  {}}

\def\la{\mathrel{\mathchoice {\vcenter{\offinterlineskip\halign{\hfil
$\displaystyle##$\hfil\cr<\cr\sim\cr}}}
{\vcenter{\offinterlineskip\halign{\hfil$\textstyle##$\hfil\cr<\cr\sim\cr}}}
{\vcenter{\offinterlineskip\halign{\hfil$\scriptstyle##$\hfil\cr<\cr\sim\cr}}}
{\vcenter{\offinterlineskip\halign{\hfil$\scriptscriptstyle##$\hfil\cr<\cr\sim\cr}}}}}

\def\Pm{\mbox{\rm Pr}_M}
\def\Rm{\mbox{\rm Re}_M}
\def\Rmcrit{\mbox{\rm Re}_{{\rm M},{\rm crit}}}
\def\Rey{\mbox{\rm Re}}

\def\half{{\textstyle{1\over2}}}

\newcommand{\ymn}[4]{, ``#4,'' {\em Monthly Notices Roy.\ Astron.\ Soc.\ }{\bf #2}, #3 (#1).}
\newcommand{\ymnS}[4]{, ``#4'' {\em Monthly Notices Roy.\ Astron.\ Soc.\ }{\bf #2}, #3 (#1).}

\newcommand{\yjfm}[4]{, ``#4,'' {\em J.\ Fluid Mech.\ }{\bf #2}, #3 (#1).}

\newcommand{\ypre}[4]{, ``#4,'' {\em Phys.\ Rev.\ E }{\bf #2}, #3 (#1).}
\newcommand{\yprl}[4]{, ``#4,'' {\em Phys.\ Rev.\ Lett.\ }{\bf #2}, #3 (#1).}

\newcommand{\yapj}[4]{, ``#4,'' {\em Astrophys.\ J.\ }{\bf #2}, #3 (#1).}

\newcommand{\yapjlS}[4]{, ``#4'' {\em Astrophys.\ J.\ Lett.\ }{\bf #2}, #3 (#1).}
\newcommand{\ypp}[4]{, ``#4,'' {\em Phys.\ Plasmas }{\bf #2}, #3 (#1).}

\newcommand{\ypf}[4]{, ``#4,'' {\em Phys.\ Fluids }{\bf #2}, #3 (#1).}

\newcommand{\yjour}[5]{, ``#5,'' {\em #2} {\bf #3}, #4 (#1).}

\newcommand{\ybookC}[3]{, {\em #2} (#3, #1),}

\begin{document}
%\preprint{NORDITA 2005-34 AP}

\title{
Hydrodynamic and hydromagnetic energy spectra from large eddy simulations}
\author{Nils Erland L.\ Haugen}
  \email{Nils.E.Haugen@sintef.no}
  \affiliation{Department of Physics, The Norwegian University of Science
      and Technology, H{\o}yskoleringen 5, N-7034 Trondheim, Norway; and
      SINTEF Energy Research, Kolbj{\o}rn Hejes Vei 1A, N-7465 Trondheim, Norway}
\author{Axel Brandenburg}
  \email{brandenb@nordita.dk}
  \affiliation{NORDITA, Blegdamsvej 17, DK-2100 Copenhagen \O, Denmark}
\date{\today,~ $ $Revision: 1.135 $ $}

\begin{abstract}
Direct and large eddy simulations of hydrodynamic and hydromagnetic
turbulence have been performed in an attempt to isolate artifacts
from real and possibly asymptotic features in the energy spectra.
It is shown that in a hydrodynamic turbulence simulation with a
Smagorinsky subgrid scale model using $512^3$ meshpoints two important
features of the $4096^3$ simulation on the Earth simulator (Kaneda et
al.\ 2003, Phys.\ Fluids {\bf 15}, L21) are reproduced: a $k^{-0.1}$
correction to the inertial range with a $k^{-5/3}$ Kolmogorov slope and the
form of the bottleneck just before the dissipative subrange.
Furthermore, it is shown that, while a Smagorinsky-type model for the
induction equation causes an artificial and unacceptable reduction
in the dynamo efficiency, hyper-resistivity yields good agreement
with direct simulations.
In the large-scale part of the inertial range, an excess of the spectral
magnetic energy over the spectral kinetic energy is confirmed.
However, a trend towards spectral equipartition at smaller scales in
the inertial range can be identified.
With magnetic fields, no explicit bottleneck effect is seen.
\end{abstract}
\maketitle

\section{Introduction}

In astrophysical magnetohydrodynamic (MHD) turbulence, e.g.\ in stars,
accretion discs, the interstellar medium, and the intergalactic medium,
the magnetic and fluid Reynolds numbers are very large.
It is therefore of great interest to perform simulations with as large 
a Reynolds number as possible. 
However, the goal of reaching astrophysical values of the magnetic
Reynolds numbers is still far out of reach.
The best we can hope for is therefore to find asymptotic trends such that one
can extrapolate into the very large Reynolds number regime.
However, even that is not really possible as the following estimate shows.
As a rule of thumb, for a purely hydrodynamical simulation 
one needs at least an order of magnitude for resolving
the dissipative subrange, one order of magnitude for the bottleneck
(a shallower spectrum just before the dissipative subrange), and 
almost an order of magnitude for the forcing to become isotropic. 
This leaves basically nothing for the inertial range--even for simulations
with  $1024^3$ meshpoints. It is therefore only with simulations
as big as $4096^3$ meshpoints \cite{Kan03} that one
begins to see an inertial range.

In MHD turbulence without imposed field, i.e.\ when the
field is self-consistently generated by dynamo action, the magnetic energy
spectrum peaks at a wavenumber that is by a certain factor larger than the
wavenumber of the kinetic energy spectrum \cite{HBD03}.
This factor has been related to the value of the critical magnetic Reynolds
number for dynamo action, $\Rmcrit$.
Specifically, $k_{\rm mag}\approx k_{\rm kin}\Rmcrit^{1/2}$ has been
suggested \cite{Sub98}, where $k_{\rm mag}$ and $k_{\rm kin}$ are the
wavenumbers of the peaks of the magnetic and kinetic energy spectra,
respectively, and $\Rmcrit\approx35$ \cite{HBD04a}.
This leads to the conclusion that in MHD turbulence
one needs an even larger Reynolds number than for purely hydrodynamical
turbulence in order to have a chance to see an inertial range.

What has been found so far is that there is a certain range,
$k_{\rm mag}\la k\la k_{\rm d}$, where the spectral magnetic energy
exceeds the spectral kinetic energy
\cite{HBD03,HBD04a}, i.e.\ there is spectral super-equipartition.
While spectral super-equipartition is not a priori implausible, it is
curious that this has not been seen in simulations with an imposed field.
Such simulations with imposed field have recently been performed
\cite{CV00,MG01,MBG03} to verify the Goldreich-Sridhar theory of MHD turbulence
\cite{GS95}.
More systematic studies of the resulting energy spectra as a function of
the imposed field strength have been carried out \cite{HB04b}, and it was found
that there is spectral equipartition only when the imposed field, $B_0$,
is of equipartition strength, i.e.\ $B_0^2\sim\mu_0\rho_0 u_{\rm rms}^2$,
where $\mu_0$ is the vacuum permeability, $\rho_0$ is the mean density,
and $u_{\rm rms}$ is the rms velocity.
If $B_0$ is larger, the magnetic spectrum is always in sub-equipartition.

The case of an imposed field is usually thought to be representative of
the conditions deep in the inertial range.
Thus, the observed super-equipartition does seem to be in conflict with
this result.
This is also supported by the well known fact
that in the solar wind, kinetic and magnetic energy spectra follow a power
law with an $-5/3$ exponent over several decades \cite{BerS04}.
In this work we want to elucidate this puzzle by comparing direct
simulations with simulations using hyperviscosity and hyperresistivity,
as well as Smagorinsky subgrid scale (SGS) modelling, in order to imitate
larger Reynolds numbers.
For recent comparisons between direct and Smagorinsky SGS simulations;
see Refs.~\cite{AMKC01,HMO01,MC02,MC02b}, where also decaying turbulence
is considered, albeit only at a resolution of $64^3$ meshpoints.
This was too small to discuss the shape of the energy spectra.
Recent simulations using hyperviscosity have shown that at large enough
resolution ($512^3$ meshpoints) the same $k^{-0.1}$ correction to the
Kolmogorov $k^{5/3}$ inertial range spectrum is seen \cite{HB04a}
as in the $4096^3$ meshpoints direct simulations of Kaneda et al.\ \cite{Kan03}.
In the present paper we compare these two simulations also with new
Smagorinsky SGS models.

We need to emphasize that throughout this paper we only deal with the case
of ``non-helical'' turbulence, i.e.\ $|\bra{\uu\cdot\nabla\times\uu}|$
is negligible (or small compared with $k_{\rm f}\bra{\uu^2}$, where
$k_{\rm f}$ is the typical forcing wavenumber).
In some sense the case of finite helicity may be regarded as more
typical \cite{MB00}.
However, with helicity there is a whole range of new problems that need
to be addressed.
For example, when using hyperresistivity the magnetic field would saturate
at an artificially enhanced value when there is helicity \cite{BS02}.
These helicity effects are now fairly well understood (see Ref.~\cite{BS05}
for a review), but in the present paper we discard these complications.

\section{Method}

We solve the compressible non-ideal MHD equations,
\begin{equation}
\frac{\DD \uu}{\DD t}=-{1\over\rho}\nabla p+ \frac{\JJ \times \BB}{\rho}
+\ff+\FF_{\rm visc},
\end{equation}
where $\DD/\DD t=\partial/\partial t+\uu\cdot\nab$ is the advective
derivative, $p$ is the pressure, $\rho$ is the density,
$\ff$ is an isotropic random nonhelical
forcing function with power in a narrow band of wavenumbers, 
$\BB$ is the magnetic field, $\JJ=\nabla\times\BB/\mu_0$ is the current
density, and $\FF_{\rm visc}$ is the viscous force (see below).
We consider an isothermal gas with constant sound speed $c_{\rm s}$,
so that the pressure is given by $p=c_{\rm s}^2\rho$ and
$\rho^{-1}\nab p=c_{\rm s}^2\nab\ln\rho$.
The density obeys the continuity equation,
\begin{equation}
\frac{\DD\ln\rho}{\DD t}=-\nab\cdot\uu.
\label{continuity}
\end{equation}
The induction equation is solved in terms of the magnetic vector
potential $\AAA$, 
\EQ
{\partial\AAA\over\partial t}=\uu\times\BB-\EE_{\rm res},
\label{dAdt}
\EN
where $\BB=\nab\times\AAA$ is the magnetic flux density,
and $\EE_{\rm res}$ is the electric field due to resistive effects (see below).

In the following, different combinations of expressions for
$\FF_{\rm visc}$ and $\EE_{\rm res}$ have been explored.
In all simulations these expressions are of the general form
\begin{equation}
\label{Fvisc+Eres}
\FF_{\rm visc}={1\over\rho}\nab\cdot\left(2\rho\nu\SSSS\right),\quad
\EE_{\rm res}=\eta\mu_0\JJ,
\end{equation}
where
\EQ
{\sf S}_{ij}=\frac{1}{2}\left({\partial u_i\over\partial x_j}
+ {\partial u_j\over\partial x_i}
-\frac{2}{3} \delta_{ij}\nab\cdot\uu\right)
\EN
is the traceless rate of strain tensor.
In a direct simulation we simply use constant values of $\nu$ and
$\eta$, i.e.\
\begin{equation}
\nu=\nu_0,\quad\eta=\eta_0\quad\mbox{(direct)}.
\end{equation}
In the case of a Smagorinsky SGS model we use
$\nu=\nu_{\rm S}$ and $\eta=\eta_{\rm S}$
(without constant contributions) where
\begin{equation}
\nu_{\rm S}=(C_{\rm K} \Delta)^2\sqrt{2\SSSS^2},\quad
\eta_{\rm S}=(C_{\rm M} \Delta)^2\sqrt{\JJ^2} \quad\mbox{(Smagorinsky)},
\end{equation}
where $C_{\rm K}$ is the Smagorinsky constant,
$C_{\rm M}$ is the magnetic Smagorinsky constant,
and $\Delta$ is the filter size, which we have set equal to the mesh size.
This version of the magnetic Smagorinsky SGS model has been
studied earlier; see, e.g., Ref.~\cite{AMKC01}.
Following our experience from earlier work \cite{HB04a}
we choose $C_{\rm K}=0.2$, but we vary the value of $C_{\rm M}$.
In simulations with hyperviscosity we replace 
\EQ
\rho\nu\SSSS\to\rho_0\nu_3\nabla^4\SSSS,\quad
\eta\JJ\to\eta_3\nabla^4\JJ \quad\mbox{(hyper)},
\EN
in \Eq{Fvisc+Eres}, and use constant coefficients, referred to as
$\nu=\nu_3$ and $\eta=\eta_3$.
Following \cite{HB04a}, we use constant {\it dynamical} hyperviscosity,
$\rho_0\nu_3=\const$,
in which case a positive viscous heating term can be defined.

In the present work we only consider cases with small Mach number.
Compressibility effects are therefore unimportant
\cite{compressibility}, and the continuity
equation \eq{continuity} can therefore be solved without additional
subgrid scale terms.
We note, however, that by defining suitable averages (Favre filtering; see
Ref.~\cite{Erle}) the continuity does formally retain its original form.
Likewise, in strongly compressible flows a turbulent bulk viscosity will
be important for smearing out shocks; see, e.g., Ref.~\cite{HBM04}.
Again, this is neglected, because we are here only interested in nearly
incompressible flows.

It is customary to quote Reynolds numbers based on the
Taylor microscale $\lambda=\sqrt{5}u_{\rm rms}/\omega_{\rm rms}$,
where $\omega_{\rm rms}$ is the rms vorticity,
and on the one-dimensional velocity dispersion $u_{\rm 1D}$, where
$u_{\rm 1D}^2=u_{\rm rms}^2/3$.
Hence, we define
the fluid and magnetic Reynolds numbers for a direct numerical simulation as
\EQ
\Rey_\lambda={u_{\rm 1D}\lambda\over\nu},\quad
\Rm={u_{\rm 1D}\lambda\over\eta},
\EN
respectively.
Their ratio is the magnetic Prandtl number,
$\Pm=\nu/\eta=\Rm/\Rey$, which is unity for all runs.
For the hyperviscous and Smagorinsky cases we define the Taylor microscale
Reynolds number, in analogy to earlier work \cite{HB04a}, as
\EQ
\Rey_\lambda
=\Rey_{\lambda,0} \left(\frac{k_{d,{\rm eff}}}{k_{\rm f}} \right)^{2/3},
\EN
where we have defined the effective Kolmogorov wavenumber,
$k_{\rm d,eff}$, whose value is found empirically by making the inertial
ranges of the spectra overlap as best as possible, and $\Rey_{\lambda,0}$
is a calibration parameter.
In an earlier paper \cite{HB04a} the calibration parameter was
found to be $\Rey_{\lambda,0}\approx 7.5$, which is also the value chosen here.

We use non-dimensional quantities by measuring length in units of $1/k_1$
(where $k_1=2\pi/L$ is the smallest wavenumber in the box of size $L$),
speed in units of the isothermal sound speed $c_{\rm s}$, density in units
of the initially uniform value $\rho=\rho_0$, and magnetic field in units of
$(\mu_0\rho_0 c_{\rm s}^2)^{1/2}$.

We use periodic boundary conditions in all three directions
for all variables. This implies that the mass in the box
is conserved, i.e.\ $\bra\rho=\rho_0$, where angular brackets denote
volume averages. We adopt a forcing function $\ff$ of the form
\EQ
\ff(\xx,t)=\Re\{N\ff_{\kk(t)}\exp[\ii\kk(t)\cdot\xx+\ii\phi(t)]\},
\EN
where $\xx$ is the position vector, and $\Re$ indicates the real part.
The wave vector $\kk(t)$ and the random phase
$-\pi<\phi(t)\le\pi$ change at every time step, so $\ff(\xx,t)$ is
$\delta$-correlated in time.
For the time-integrated forcing function to be independent
of the length of the time step $\delta t$, the normalization factor $N$
has to be proportional to $\delta t^{-1/2}$.
On dimensional grounds it is chosen to be
$N=f_0 c_{\rm s}(|\kk|c_{\rm s}/\delta t)^{1/2}$, where $f_0$ is a
non-dimensional forcing amplitude.
The value of the coefficient $f_0$ is chosen such that the maximum Mach
number stays below about 0.2.
Empirically, this is achieved by taking $f_0=0.02$ for all runs discussed below.

At each timestep we select randomly one of many possible wave vectors
in a certain range
around a given forcing wavenumber.
The average wavenumber is referred to as $k_{\rm f}$.
We force the system with nonhelical transversal waves,
\EQ
\ff_{\kk}=\left(\kk\times\eee\right)/\sqrt{\kk^2-(\kk\cdot\eee)^2},
\label{nohel_forcing}
\EN
where $\eee$ is an arbitrary unit vector that is real and not aligned
with $\kk$; note that $|\ff_{\kk}|^2=1$.
For all simulations we use the {\sc Pencil Code} 
(\url{http://www.nordita.dk/software/pencil-code}) which
is a grid based high order code (sixth order in space and third order
in time) for solving the compressible hydromagnetic equations.

\section{Results}

In an earlier paper \cite{HB04a} we have shown that hyperviscosity,
although it does cause an artificially enhanced bottleneck effect
in purely hydrodynamic turbulence,
it does not affect the inertial range if the resolution is large enough.
Instead, hyperviscous simulations with $512^3$ meshpoints reproduce the
$k^{-0.1}$ correction with wavenumber $k$.
This was first found by Kaneda et al.\ \cite{Kan03}.
We begin by comparing these results with simulations where
Smagorinsky SGS viscosity is used.

\subsection{Hydrodynamic turbulence}
\label{Hydro}

\begin{figure}\centering\includegraphics[width=0.50\textwidth]
{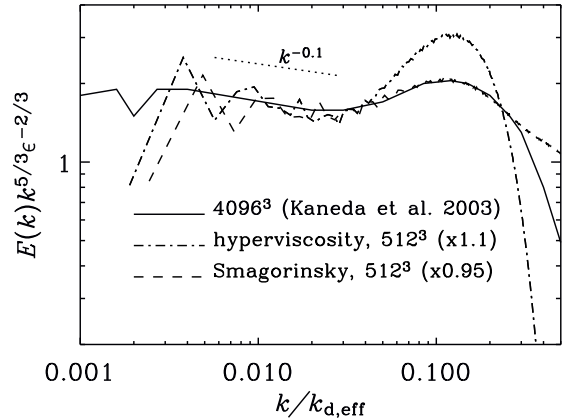}
\caption{Comparison of energy spectra of the $4096^3$ meshpoints run 
\cite{Kan03} (solid line) and $512^3$ meshpoints runs with 
hyperviscosity (dash-dotted line) and Smagorinsky viscosity (dashed line).
(In the hyperviscous simulation we use $\nu=\nu_3=5 \times 10^{-13}$.)
The Taylor microscale Reynolds number of the Kaneda simulation is 1201,
while the hyperviscous simulation of Ref.~\cite{HB04a} has an approximate
Taylor microscale Reynolds number of $340<\mbox{Re}_\lambda<730$. 
For the Smagorinsky simulation the value of $\mbox{Re}_\lambda$ is
slightly smaller.
\label{kan_hyp_smag}}
\end{figure}

In \Fig{kan_hyp_smag} we compare kinetic energy spectra of runs using
ordinary viscosity ($4096^3$ meshpoints, solid line) by
Kaneda et al.\ \cite{Kan03} 
with runs using Smagorinsky viscosity ($512^3$ meshpoints, dashed line)
and runs using hyperviscosity ($512^3$ meshpoints, dash-dotted line).
Since the simulation with $4096^3$ meshpoints and ordinary viscosity is
the largest direct simulation to date, we use it as our benchmark.
The spectra for the runs with hyperviscosity and Smagorinsky viscosity
have been scaled by empirically determined factors 1.1 and 0.95,
respectively, so as to make the spectra overlap within the inertial range.
However, these scaling factors are still well within the range over
which the spectra fluctuate in time.

We see that at all scales (including those of the bottleneck)
the simulation with Smagorinsky SGS modeling
is surprisingly similar to the benchmark result.
Furthermore we see that at large scales and in the inertial range
the run with hyperviscosity agrees well with the benchmark result.
The bottleneck is however greatly exaggerated in height, even though the 
width is the same \cite{HB04a}. 

Most important is perhaps the $k^{-0.1}$ correction to the usual
$k^{-5/3}$ inertial range scaling.
The same correction is seen in all three simulations.
The $k^{-0.1}$ correction is stronger than the usual intermittency
correction predicted by the She-Leveque model \cite{SL94}, which
would only predict a $k^{-0.03}$ correction.
This strong correction may be an artifact of the absence of a well
resolved subinertial range \cite{Davidson04}.
This would be in some ways just opposite to the emergence of a
shallower spectrum near the dissipative cutoff wavenumber if the
dissipative subrange is not well resolved \cite{Falk}.

The only major discrepancy between the Smagorinsky and direct simulations
is the lack of a sharp decline of the spectral energy toward the right of
the bottleneck.
In order to understand this difference, we must first of all recall that
our Smagorinsky simulation did not have any explicit (constant) component
at all ($\nu_0=0$).
Therefore, if the Smagorinsky model was a perfect subgrid scale model,
it should represent the infinite Reynolds number case.
The bottleneck would then be far to the right and outside the graph,
so one should only have a pure Kolmogorov spectrum.
The reason for the bottleneck in the Smagorinsky case is therefore
related to the fact that we are still working here with an ordinary
diffusion operator using just a variable viscosity coefficient.
Therefore, the standard explanation for the bottleneck still applies;
it is caused by strongly nonlocal interactions in wavenumber space,
corresponding to wave vectors forming strongly elongated triangles.
Close to the viscous cutoff wavenumber, these interactions prevent
the disposal of energy from the end of the inertial range, which then
causes the pileup of energy near the dissipation wavenumber \cite{Falk}.
The same argument also applies to the current case of Smagorinsky viscosity.
In conclusion, the reason for the discrepancy between direct and Smagorinsky
simulations to the right of the bottleneck is that the Smagorinsky model
tries to maintain pure Kolmogorov scaling everywhere, but fails to do so
just before the cutoff wavenumber imposed by the finite mesh resolution.

\begingroup
\begin{table}[t!]
\centering
\caption{
Summary of the most important runs.
The meaning of entries in the columns for $\nu$ and $\eta$ depends on
the entry for `Method', as explained in the text.
In the Smagorinsky cases ordinary viscosity is neglected, i.e.\ $\nu=0$.
Except for Method~O, the resulting values of $k_{d,{\rm eff}}$, and hence
also of $\Rey_{\lambda}$, are uncertain within $\sim40\%$
}
\label{Tsum}
\begin{ruledtabular}
\begin{tabular}{lcccccl}
Run &  Res.    & Method & $\nu$             & $\eta$& $k_{d,{\rm eff}}$ &$\Rey_{\lambda}$\\
\hline
A   & $1024^3$ & O      & $8\times 10^{-5}$ & $8\times 10^{-5}$  & 143 &
200\footnote{Note that in Ref.~\cite{DHYB03} the value of $\Rey_{\lambda}$
was based on the 3-dimensional velocity dispersion, so the non-magnetic
equivalent of Run~A was quoted with $\Rey_{\lambda}=350$.
}\\
B1  & $ 128^3$ & II     &     0             & $1\times 10^{-9}$  & $180$ & $180$\\
B2  & $ 256^3$ & II     &     0             & $3\times 10^{-11}$ & $330$ & $270$\\
B3  & $ 512^3$ & II     &     0             & $5\times 10^{-13}$ & $700$ & $450$\\
C1  & $ 128^3$ & III    & $1\times 10^{-9}$ & $1\times 10^{-9}$  & $180$ & $180$\\
C2  & $ 256^3$ & III    & $3\times 10^{-11}$& $3\times 10^{-11}$ & $330$ & $270$\\
C3  & $ 512^3$ & III    & $5\times 10^{-13}$& $5\times 10^{-13}$ & $700$ & $450$\\
\end{tabular}
\end{ruledtabular}
\end{table}
\endgroup

\subsection{Hydromagnetic turbulence}

For the MHD case we use a $1024^3$ meshpoints simulation with ordinary 
viscosity as our benchmark \cite{HBD03}. 
We compare with the SGS model where Smagorinsky schemes
are used both for the velocity and the magnetic fields.
In the following we refer to this as Method~I.
We also compare with cases where we use hyperresistivity.
In the momentum equation we use either the usual Smagorinsky SGS
model, which is referred to as Method~II,
or we use hyperviscosity (Method~III).
The results of these three methods are compared with those of
direct simulations (Method~O).
In summary, the different methods considered here are
\begin{eqnarray}
\begin{array}{lrllll}
\mbox{Method} & \mbox{I:}  & \nu_{\rm S} & \mbox{and} & \eta_{\rm S} & \mbox{(full Smagorinsky)}, \nonumber \\
\mbox{Method} & \mbox{II:} & \nu_{\rm S} & \mbox{and} & \eta_3 & \mbox{(Smagorinsky/hyper)}, \nonumber \\
\mbox{Method} & \mbox{III:}& \nu_3       & \mbox{and} & \eta_3 & \mbox{(full hyper)}, \nonumber \\
\mbox{Method} & \mbox{O:}  & \nu_0       & \mbox{and} & \eta_0 & \mbox{(benchmark)}.
\end{array}
\end{eqnarray}
We have listed the relevant runs in \Tab{Tsum}.

\begin{figure}\centering\includegraphics[width=0.50\textwidth]
{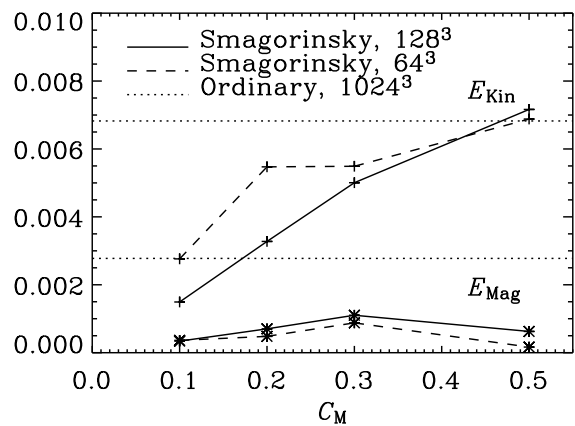}\caption{
Total magnetic and kinetic energies for runs with
$128^3$ (solid line) and $64^3$ (dashed line) meshpoints and 
Smagorinsky diffusion and resistivity (Method~I)
compared with a direct simulation with $1024^3$ meshpoints
(Method~O, horizontal dotted lines). 
Note the lack of convergence for any value of $C_{\rm M}$.
}\label{compare_smagorinsky128}\end{figure}

In \Fig{compare_smagorinsky128} we show that the agreement between
the results of Method~I and the benchmark is poor.
The dynamo-generated magnetic energy remains far below the benchmark target.
The largest value of the magnetic energy is reached for $C_{\rm M}=0.3$,
but even then it is only about $30\%$ of the target value.

\begin{figure}\centering\includegraphics[width=0.50\textwidth]
{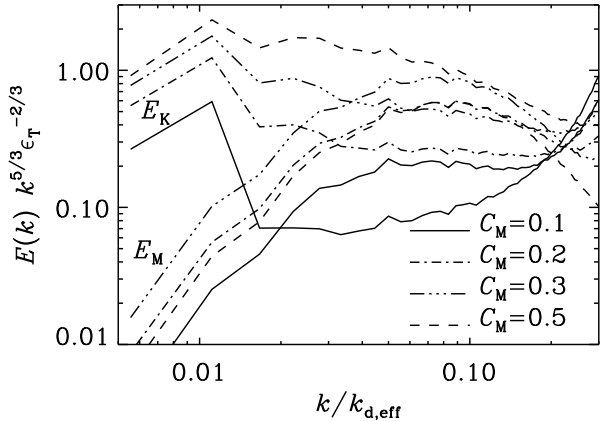}\caption{
Comparison of magnetic and kinetic energy spectra of runs using Method~I
with $128^3$ meshpoints and various values of $C_{\rm M}$.
}\label{smag_smag_comp}\end{figure}

In order to understand the reason for the poor performance of the
magnetic Smagorinsky model (Method~I) we plot in \Fig{smag_smag_comp}
kinetic and magnetic energy spectra for various values of $C_{\rm M}$.
Clearly, for $C_{\rm M}\le0.3$ both kinetic and magnetic spectra
diverge toward large wavenumbers.
This shows that this model becomes unphysical and cannot be used for
too small values of $C_{\rm M}$.
For $C_{\rm M}=0.5$, on the other hand, magnetic and kinetic spectra
fall off at large wavenumbers.
However, the effective resistivity of the magnetic Smagorinsky scheme
is apparently too large for $C_{\rm M}=0.5$, so that the dynamo
is suppressed.
The poor performance of this model is not too surprising if one recalls
that it is a rather crude method in that it deals with the small scales
only in a diffusive manner.
We also note that the Smagorinsky SGS model has, to our knowledge,
never before been tested in the context of dynamo action.
We conclude that using the Smagorinsky SGS model for the magnetic
field does not give satisfactory results.
Therefore, from now on, we discard it as inappropriate for our purpose.

\begin{figure}\centering\includegraphics[width=0.50\textwidth]
{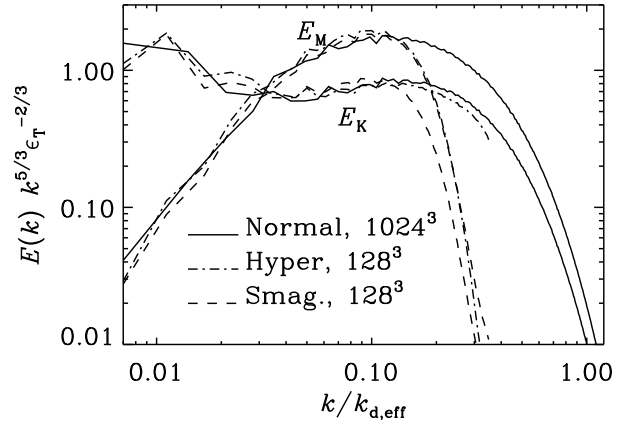}\caption{
Comparison of magnetic and kinetic 
energy spectra of runs with $1024^3$ meshpoints and normal
diffusion (Run A, solid line)
with $128^3$ meshpoints and hyperdiffusion (Run C1,
dash-dotted line), and with 
$128^3$ meshpoints and Smagorinsky viscosity and hyperresistivity 
(Run B1, dashed line).
Note that both the magnetic and kinetic energy spectra for the three runs 
are very similar for $k/k_{\rm d}<0.1$.
}\label{normal_comp}\end{figure}

We see from \Fig{normal_comp} that the compensated spectra
with only $128^3$ meshpoints, using Methods~II and III,
match the benchmark quite well at all scales down to the 
dissipative scale.
We have compensated the energy spectra by
$k^{5/3}\epsilon_{\rm T}^{-2/3}$, such that a
Kolmogorov-like spectrum would appear flat.
Here $\epsilon_{\rm T}=\epsilon_{\rm K}+\epsilon_{\rm M}$, where 
$\epsilon_{\rm K}$ and $\epsilon_{\rm M}$ are the kinetic and magnetic 
dissipation rates, respectively.
The kinetic energy spectrum of the $1024^3$ run has however been multiplied 
by $1.3$ in order to make all spectra overlap. We believe the shift is
due to the fact that the $1024^3$ run has not been run for very long, and the
average dissipation rate, $\epsilon_{\rm T}$, has not yet fully converged,
even though the slope converges generally much quicker.
 From the general agreement between the three runs shown in \Fig{normal_comp}
we conclude that, for our purpose, Methods~II and III give useful results.

\begin{figure}\centering\includegraphics[width=0.50\textwidth]
{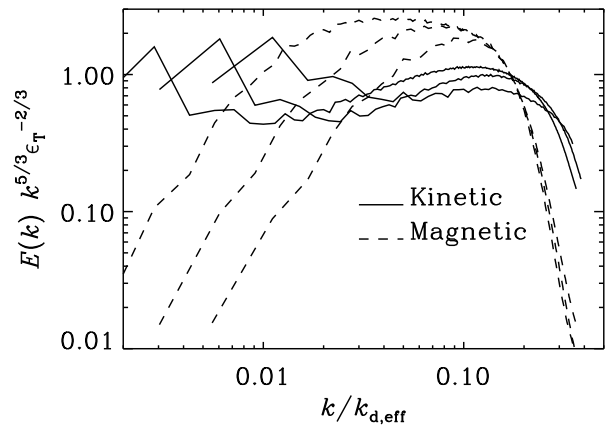}\caption{
Magnetic and kinetic energy spectra for runs with
$128^3$ (Run B1), $256^3$ (Run B2) and $512^3$ (Run B3) meshpoints 
where all of them use Smagorinsky viscosity and
hyperresistivity (Method II).
Note the approach of the kinetic energy spectra towards the
magnetic energy spectra at a point that is well before entering
the bottleneck and the dissipative subrange.
}\label{smag_comp}\end{figure}

\begin{figure}\centering\includegraphics[width=0.50\textwidth]
{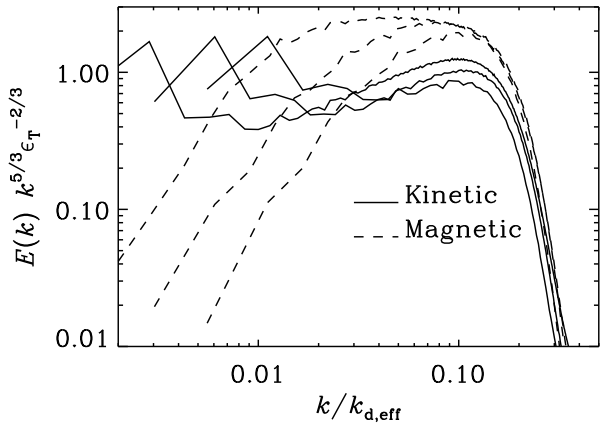}\caption{
Magnetic and kinetic energy spectra for runs with
$128^3$ (Run C1), $256^3$ (Run C2) and $512^3$ (Run C3) meshpoints 
where all of them use hyperviscosity and
hyperresistivity (Method III).
}\label{hyper_comp}\end{figure}

\begin{figure}\centering\includegraphics[width=0.50\textwidth]
{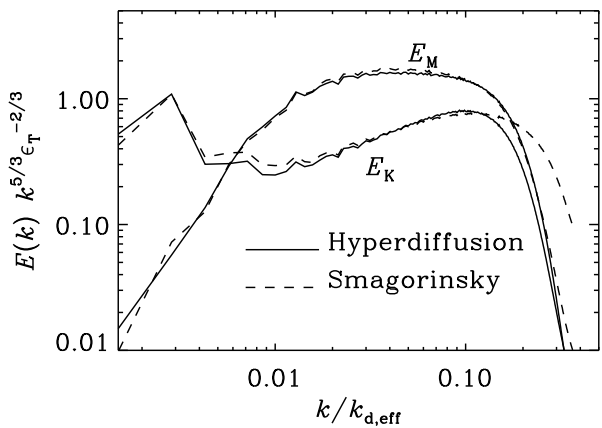}\caption{
Magnetic and kinetic energy spectra for runs with $512^3$ 
meshpoints and hyperviscosity and hyperresistivity (Run C3, solid line) and 
Smagorinsky viscosity and hyperresistivity (Run B3, dashed line).
Note the mutual approach of kinetic and magnetic energy spectra
before entering the dissipative subrange.
}\label{hyper_512}\end{figure}

In \Fig{smag_comp} we compare compensated spectra for three 
simulations which all use Smagorinsky viscosity and hyperresistivity,
but have different 
Reynolds numbers.
We see that, unlike the purely hydrodynamic case, the dissipative
subranges do not collapse onto the same functional form for different
Reynolds numbers.
On the other hand, for purely hydrodynamical simulations \cite{HB04a}
the dissipative subranges collapse
very well onto the same functional form and the inertial range simply
becomes longer for larger Reynolds numbers.
Furthermore, in Fig~1 of Ref.~\cite{HB04a} we see that the bottleneck
is similar and constant for all Reynolds numbers.
Again, in the MHD simulation we see nothing similar. 

In \Fig{hyper_comp} we have shown the same as in \Fig{smag_comp}, but
using hyperviscosity instead of Smagorinsky viscosity. 
We clearly see that the tendency is the same in both figures. 
Since the bottleneck effect is quite different for pure hydrodynamical 
simulations with
Smagorinsky viscosity and with hyperviscosity (see \Fig{kan_hyp_smag}),
it is reasonable to assume that the tendency we see is robust and not
due to the specific modeling applied, but that it is a physical effect.

Finally, we compare in \Fig{hyper_512} spectra of Smagorinsky and hyperviscous
simulations using the highest available resolution of $512^3$ meshpoints.
Again, note that the spectra for hyperviscous simulations and those with
Smagorinsky SGS modeling are almost identical.
Furthermore, there is no range where both kinetic and magnetic 
energy spectra are parallel.
Together with the results of \Figs{smag_comp}{hyper_comp}
we therefore conclude that we have not yet reached
Reynolds numbers large enough to show an inertial range. 

\section{Speculations on asymptotics}

The direct MHD simulations of Ref.~\cite{HBD03} have suggested the presence
of a super-equipartition range where $E_{\rm M}(k)\sim2.5E_{\rm K}(k)$.
However, the spectra still showed some weak bending, indicating that
a proper inertial range has not been reached even at a resolution of
$1024^3$ meshpoints \cite{Scheko04}.
The present SGS models reproduce the spectral super-equipartition
of magnetic over kinetic spectral energy (\Fig{hyper_512}), but they also
show now more clearly that the two spectra are not parallel to each other.
Instead, they approach each other in such a way that the compensated
kinetic energy spectrum shows a strong uprise.

\begin{figure}\centering\includegraphics[width=0.5\textwidth]
{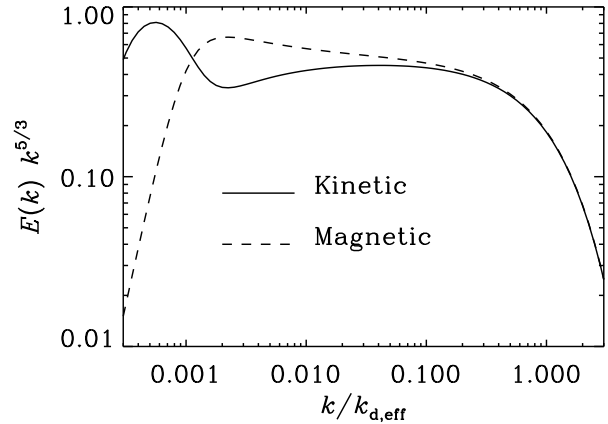}\caption{
Sketch of kinetic and magnetic energy spectra, following the
M\"uller and Grappin phenomenology.
Note the slight super-equipartition just to the right
of the peak of $E_{\rm M}(k)$ and the asymptotic
equipartition for large wavenumbers.
}\label{MullerGrappin}\end{figure}

One might argue that the uprise at the end of the compensated kinetic spectrum is 
just a strong bottleneck.
This is however unlikely since both SGS models give the same uprise,
even though in purely hydrodynamic turbulence the hyperviscosity model
is known to produce a much higher bottleneck than the Smagorinsky model
(\Sec{Hydro}).
Furthermore, in hydrodynamic turbulence the width of the bottleneck is
independent of Reynolds number, whereas in the present case it appears
to become wider with increasing Reynolds number.
This suggests that the uprise in the MHD case is a true
large scale feature of the spectrum, and independent of the dissipative
subrange.

\begin{figure}\centering\includegraphics[width=0.5\textwidth]
{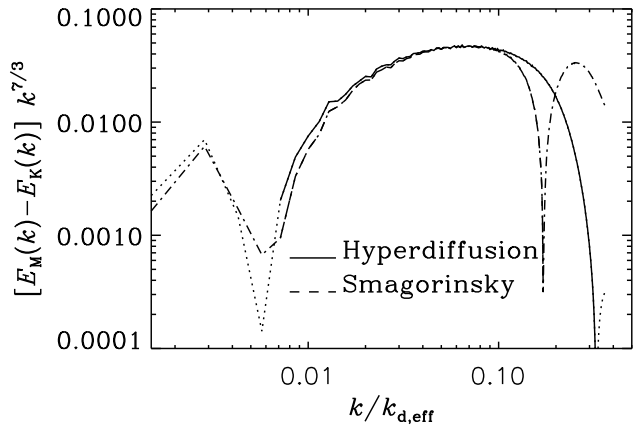}\caption{
Residual spectrum, $E_{\rm R}=E_{\rm M}-E_{\rm K}$, compensated by
$k^{7/3}$, for the same runs shown in \Fig{hyper_512}.
Negative values of $E_{\rm M}-E_{\rm K}$ are indicated by dotted and
dash-dotted lines for hyperdiffusion and Smagorinsky runs, respectively.
}\label{hyper_smag_residual}\end{figure}

Next, we recall that in
simulations with an imposed magnetic field, the magnetic and kinetic 
energy spectra are found to be in approximate equipartition only when
the field strength is of the order of $B_{\rm eq}$ \cite{HB04b}.
Such simulations are thought to be representative of the small scale
end of the inertial range of any MHD simulation, even if the field is
generated by a small scale dynamo as in the present case.
Assuming that this interpretation is correct, it would support our
previous suggestion that the spectral super-equipartition was only
a non-asymptotic feature, confined to the large scales, and not a
true inertial range feature.
We are therefore led to believe that for much larger Reynolds numbers
the kinetic and magnetic energy spectra might converge.
Qualitatively, this can be reproduced by the phenomenology proposed
recently by M\"uller and Grappin \cite{MG04,MG05}.
According to their theory, the total energy $E_{\rm T}=E_{\rm M}+E_{\rm K}$
follows still the expected $k^{-5/3}$ spectrum, while the residual energy
$E_{\rm R}=|E_{\rm M}-E_{\rm K}|$ follows a $k^{-7/3}$ spectrum.
In \Fig{MullerGrappin} we produce such an example with
\EQ
E_{\rm T}(k)=k^{-5/3}e^{-k/k_{\rm d}},\quad
E_{\rm R}(k)=ak^{-7/3}e^{-k/k_{\rm d}},
\EN
for $k\ge1$ (in arbitrary units).
Using the fact that in the inertial range $E_{\rm M}$ exceeds $E_{\rm K}$
by about a factor $a=2$, we reconstruct $E_{\rm M}$ and introduce an
additional $k^2$ subinertial range, so we write
\EQ
E_{\rm M}(k)=\half[E_{\rm T}(k)+E_{\rm R}(k)]/[1+(k/k_{\rm M})^{-11/3}],
\EN
with $k_{\rm M}=5$.
The kinetic energy obtained by assuming that the total energy is constant is
\EQ
E_{\rm K}(k)=[E_{\rm T}(k)-E_{\rm M}(k)]/[1+(k/k_{\rm K})^{-11/3}],
\EN
where we have included a different subinertial range below $k_{\rm K}=1.5$.
The resulting spectra shown in \Fig{MullerGrappin} reproduce surprisingly
well the basic features suggested by our SGS simulations of \Fig{hyper_512}.

In order to see how well our simulations reproduce the anticipated $k^{-7/3}$
scaling of the residual spectrum, we plot in \Fig{hyper_smag_residual} the
appropriately compensated $E_{\rm R}$ spectrum.
Clearly, the residual spectrum is still curved, but it remains reasonably
straight within about half an order of magnitude in wavenumbers.

\section{Conclusions}

The results of subgrid scale models should always be taken with great care.
Even if their results can be trusted in one case (e.g.\ in the case without
magnetic fields), they may not give reliable results in another case
(e.g.\ in the presence of magnetic fields and dynamo action).
However, once we begin to see detailed agreement between SGS models and
direct simulations, it may be possible to use this agreement to justify
the use of the SGS model in more extreme parameter regimes that are
currently inaccessible to direct simulations.

In the present work we have shown that the Smagorinsky SGS model with a
resolution of $512^3$ meshpoints is able to reproduce the hydrodynamic
turbulence spectra of a direct simulation at an almost 10 times larger
resolution (\Fig{compare_smagorinsky128}).
On the other hand, an extension of this model to the MHD case with
dynamo action leads to obvious problems (the intensity of the dynamo is
artificially suppressed).
However, using hyperresistivity instead of a Smagorinsky-type SGS model
leads to fair agreement between the $128^3$ SGS simulation and the nearly
10 times larger direct simulation (\Fig{normal_comp}).
Thus, having validated the SGS model at $128^3$ meshpoints, we may be
justified in proceeding further to a resolution of $512^3$ meshpoints
(\Fig{hyper_512}).
Here, a new and yet unconfirmed feature arises: a {\it tendency} towards
spectral equipartition.
This, together with the knowledge that there is spectral equipartition
with imposed fields of equipartition strength \cite{HB04b}, suggests
a spectrum that might look like what is shown in \Fig{MullerGrappin}.

Obviously, we will not be able to verify this result in the immediate
future.
Although it may soon be possible to obtain the resources necessary to do
a $4096^3$ MHD simulation to validate the results of \Fig{hyper_512},
yet another order of magnitude in improved resolution will be necessary
to test the hypothesis sketched in \Fig{MullerGrappin}.
Our results may therefore serve as a justification for using future
computing resources for this type of problem.

\section*{Acknowledgements}

We thank the three anonymous referees for making useful suggestions that
have improved the presentation of the paper.
This work has been supported in part by a David Crighton Fellowship
and by the Isaac Newton Institute in Cambridge, where part of this work
has been carried out.
We thank the Danish Center
for Scientific Computing for granting time on the Horseshoe cluster,
and the Norwegian High Performance Computing Consortium (NOTUR)
for granting time on the parallel computers in 
Trondheim (Gridur/Embla) and Bergen (Fire).

%r e f

\label{lastpage}

\begin{thebibliography}{}
\bibitem{compressibility}
For this type of a weakly compressible simulation, we find that the
energies of solenoidal and potential components of the flow have a ratio
$E_{\rm pot}/E_{\rm sol} \approx 10^{-4}\mbox{--}10^{-2}$ at most scales;
only towards the Nyquist frequency the ratio increases to about $0.1$.
It is thus reasonable to assume that compressibility is irrelevant for the
results presented here.
\end{thebibliography}
\end{document}